\documentclass[aps, prd, floatfix, nofootinbib, showpacs, twocolumn, superscriptaddress]{revtex4-1}

\usepackage{amsmath}
\usepackage{amssymb}
\usepackage{graphicx}
\usepackage{color}
\usepackage{xspace}

\usepackage[mathscr,scaled=1.15]{urwchancal}
\DeclareFontFamily{OT1}{pzc}{}
\DeclareFontShape{OT1}{pzc}{m}{it}%
{<-> s * [1.15] pzcmi7t}{}
\DeclareMathAlphabet{\mathpzc}{OT1}{pzc}{m}{it}

\definecolor{purple}{rgb}{0.5,0,0.5}
\definecolor{blue}{rgb}{0.0,0,0.9}
\definecolor{prdblue}{rgb}{0.133,0.118,0.498}
\usepackage[colorlinks=true, pdfstartview=FitV, linkcolor=prdblue, citecolor= prdblue, urlcolor=prdblue]{hyperref}

\begin{document}

\title{Proton tensor charges from a Poincar\'e-covariant Faddeev equation}

\author{Qing-Wu~Wang}
\affiliation{Department of Physics, Sichuan University, Chengdu 610064, P.R. China}

\author{Si-Xue~Qin}
\email[]{sqin@cqu.edu.cn}
\affiliation{Department of Physics, Chongqing University, Chongqing 401331, P.R. China}

\author{Craig D.~Roberts}
\affiliation{Physics Division, Argonne National Laboratory, Argonne,
Illinois 60439, USA}

\author{Sebastian M. Schmidt}
\affiliation{
Institute for Advanced Simulation, Forschungszentrum J\"ulich and JARA, D-52425 J\"ulich, Germany}

\date{31 May 2018}

\begin{abstract}
The proton's tensor charges are calculated at leading order in a symmetry-preserving truncation of all matter-sector equations relevant to the associated bound-state and scattering problems.  In particular, the nucleon three-body bound-state equation is solved without using a diquark approximation of the two-body scattering kernel.  The computed charges are similar to those obtained in contemporary simulations of lattice-regularised quantum chromodynamics, an outcome which increases the tension between theory and phenomenology.  Curiously, the theoretical calculations produce a value of the scale-invariant ratio $(-\delta_T d/\delta_T u)$ which matches that obtained in simple quark models, even though the individual charges are themselves different.  The proton's tensor charges can be used to constrain extensions of the Standard Model using empirical limits on nucleon electric dipole moments.
\end{abstract}

\maketitle


\section{Introduction}
\begin{figure}[t!]
\begin{center}
\includegraphics[clip,width=0.35\textwidth]{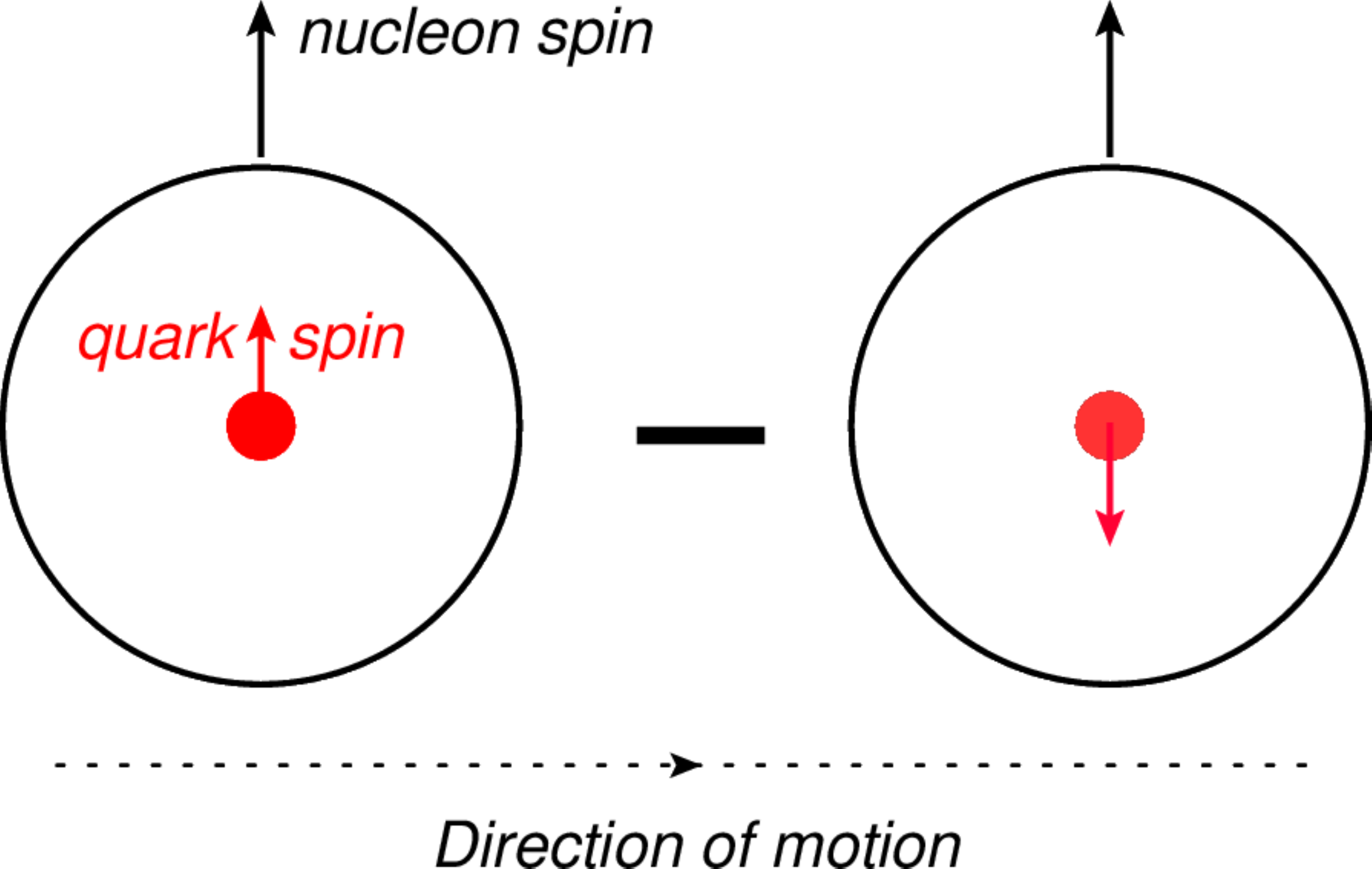}
\caption{\label{figTensorCharge} The tensor charge, Eq.\,\eqref{DefineTensorCharge}, measures the net light-front distribution of transversely polarised quarks inside a transversely polarized proton.}
\end{center}
\end{figure}

New generation experiments \cite{Dudek:2012vr, Gao:2010av, Avakian:2014aba, Gao:2017ade} aim to obtain data that can be used to determine the proton's transverse momentum dependent parton distribution functions (TMDs) \cite{Ralston:1979ys, Sivers:1989cc, Kotzinian:1994dv, Mulders:1995dh, Collins:2003fm, Belitsky:2003nz, Bacchetta:2006tn}.  At leading-twist, three distinct TMDs are nonzero in the collinear limit, \emph{i.e}.\ in the absence of parton transverse momentum within the target, $k_\perp = 0$: the unpolarized $(f_1)$, helicity $(g_{1L})$ and transversity $(h_{1T})$ distributions.  The last of these may be used to express the proton's tensor charges ($q=u,d,\ldots$)
\begin{equation}
\label{DefineTensorCharge}
\delta_T q = \int_{-1}^1 dx\, h^q_{1T}(x) = \int_0^1 dx\, \left[ h_{1T}^q(x) - h_{1T}^{\bar q}(x)\right]\,,
\end{equation}
which, as illustrated in Fig.\,\ref{figTensorCharge}, measures the light-front number-density of quarks with transverse polarisation parallel to that of the proton minus that of quarks with antiparallel polarisation; namely, it measures any bias in quark transverse polarisation induced by a polarisation of the parent proton.

The tensor charges $\delta_T q$ are close analogues of the nucleon flavour-separated axial-charges, which measure the difference between the light-front number-density of quarks with helicity parallel to that of the proton and the density of quarks with helicity antiparallel.  In nonrelativistic systems, the helicity and transversity distributions are identical because boosts and rotations commute with the Hamiltonian.  This connection highlights the fundamental nature of tensor charges: they are a defining property of the nucleon and may be judged to measure, \emph{inter alia}, the importance of Poincar\'e-covariance in treatments of the nucleon bound state.

One can also compute the tensor charge associated with a given quark in the proton via the matrix element
\begin{align}
\label{tcd}
\langle P(k,\sigma)|\bar q\sigma_{\mu\nu}q|P(k,\sigma)\rangle=\delta_T q\,\bar
{\mathpzc u}(k,\sigma)\sigma_{\mu\nu}{\mathpzc u}(k,\sigma) \,,
\end{align}
where ${\mathpzc u}(k,\sigma)$ is a spinor and $|P(k,\sigma)\rangle$ is a state vector describing a proton with momentum $k$ and spin $\sigma$.  Importantly, the tensor charge is a scale-dependent quantity and this must be borne in mind when comparing results from
different calculations.  Naturally, in the isospin symmetric limit:
\begin{equation}
\delta_T u :=\delta_T^p u = \delta_T^n d\,,\;
\delta_T d := \delta_T^p d = \delta_T^n u\,;
\end{equation}
and using $\delta_T u$, $\delta_T d$, the isoscalar and isovector tensor charges are readily computed:
\begin{equation}
  g_T^{(0)} = \delta_T u + \delta_T d\,, \;
  g_T^{(1)} = \delta_T u - \delta_T d\,.
\end{equation}
The value of the last of these, $g_T^{(1)}$, bears comparison with the nucleon axial charge $g_A = 1.276$ \cite{Mendenhall:2012tz, Mund:2012fq}.

Apart from the hadron physics interest, the value of the nucleon tensor charges can also be used to constrain new physics.  This is because in typical extensions of the Standard Model (SM), quarks acquire an electric dipole moment (EDM) \cite{Pospelov:2005pr, RamseyMusolf:2006vr}, \emph{viz}.\ an interaction with the photon that proceeds via a current of the form:
\begin{equation}
\label{EDMoperator}
\tilde d_q \, q \gamma_5 \sigma_{\mu\nu} q\,,
\end{equation}
where $\tilde d_q$ is a quark EDM of unknown magnitude.  Despite the fact that $\tilde d_q \neq 0$ violates both parity and time-reversal invariance, this does not itself produce a conflict with the SM.  The challenge to new physics is found in the fact that the first nonzero SM contribution to a quark's EDM appears at third order and involves a gluon radiative correction, so that $\tilde d_q^{\rm SM} \lesssim 10^{-34}e\cdot$cm \cite{He:1990qa}, a value so small that SM-extensions are very tightly constrained.

Consider now the EDM of a proton containing quarks which interact via Eq.\,\eqref{EDMoperator}:
\begin{subequations}
\begin{align}
\langle P(k,\sigma) | {\mathcal J}_{\mu\nu}^\text{EDM} & | P(k,\sigma)\rangle
= \tilde d_p\, \bar {\mathpzc u}(k,\sigma)\gamma_5\sigma_{\mu\nu} {\mathpzc u}(k,\sigma)\,,\\
 \mathcal J_{\mu\nu}^\text{EDM}= \tilde d_u\, \bar u&(x)  \gamma_5\sigma_{\mu\nu}u(x) + \tilde d_d\,\bar d(x)\gamma_5\sigma_{\mu\nu}d(x)\,.
\end{align}
\end{subequations}
Using a Dirac-matrix identity: $
\gamma_5\sigma_{\mu\nu} = \tfrac{1}{2} \varepsilon_{\mu\nu\alpha\beta} \sigma_{\alpha\beta}$,
\begin{equation}
\mathcal J_{\mu\nu}^\text{EDM}
= \tfrac{1}{2} \varepsilon_{\mu\nu\alpha\beta}
\left[ \tilde d_u\,\bar u \sigma_{\alpha\beta}u  + \tilde d_d\,\bar d\sigma_{\alpha\beta}d\right]\,.
\end{equation}
Hence,
\begin{subequations}
\begin{align}
&\langle P(k,\sigma) | \mathcal J_{\mu\nu}^\text{EDM} | P(k,\sigma)\rangle \nonumber\\
&= \tfrac{1}{2} \varepsilon_{\mu\nu\alpha\beta}
\left[
\tilde d_u \, \delta_T u\,
+ \tilde d_d \, \delta_T d\,
\right]\bar {\mathpzc u}(k,\sigma)\sigma_{\alpha\beta}{\mathpzc u}(k,\sigma) \\
&= \left[
\tilde d_u \, \delta_T u\,
+ \tilde d_d \, \delta_T d\,
\right]\bar {\mathpzc u}(k,\sigma)\gamma_5\sigma_{\mu\nu}{\mathpzc u}(k,\sigma) \,;
\end{align}
\end{subequations}
namely, the quark-EDM contribution to a proton's EDM is determined once the proton's tensor charges are known:
\begin{subequations}
\label{EDMeqs}
\begin{equation}
\tilde d_p = \tilde d_u \, \delta_T u\, + \tilde d_d \, \delta_T d\,.
\end{equation}
With emerging techniques, it is becoming possible to place competitive upper-limits on the proton's EDM using storage rings in which polarized particles are exposed to an electric field \cite{Pretz:2013us}.

Using isospin symmetry,
\begin{equation}
\tilde d_n = \tilde d_u \, \delta_T d\, + \tilde d_d \, \delta_T u\,.
\end{equation}
\end{subequations}
Empirically \cite{Schmidt-Wellenburg:2016nfv}: $\tilde d_n < 3 \times 10^{-26}e\cdot$cm.

Given their importance, the proton's tensor charges have been computed using a variety of methods, with progress recently using lattice-regularised QCD (lQCD) \cite{Bhattacharya:2015esa, Bhattacharya:2016zcn, Alexandrou:2017qyt, Bali:2014nma}.
Continuum methods have also been employed \cite{He:1994gz, Hecht:2001ry, Yamanaka:2013zoa, Pitschmann:2014jxa, Xu:2015kta}; and herein we report the most refined such calculation to date, using a symmetry-preserving approach to the continuum bound-state problem in QCD \cite{Roberts:1994dr, Chang:2011vu, Bashir:2012fs, Roberts:2015lja, Horn:2016rip, Eichmann:2016yit}.

We represent the proton by the solution of a three-body analogue of the Bethe-Salpeter equation, commonly described as a Poincar\'e-covariant Faddeev equation.  This approach to baryons was introduced in Refs.\,\cite{Cahill:1988dx, Burden:1988dt, Reinhardt:1989rw, Efimov:1990uz}, which capitalised on the role of diquark correlations in order to simplify the problem \cite{Cahill:1987qr, Maris:2002yu, Maris:2004bp, Segovia:2015hra, Segovia:2015ufa, Segovia:2016zyc, Chen:2017pse, Roberts:2018hpf}; but we adapt the formulation in Refs.\,\cite{Eichmann:2009qa, Eichmann:2011vu} and solve the three-valence-body problem directly, under the assumption that two-body interactions dominate in forming a baryon bound-state.  This means that we also solve the dressed-quark gap equation and inhomogeneous Bethe-Salpeter equation for the quark tensor vertex using the same interaction.

We describe the three-body bound-state equation and the character of its solution for the nucleon in Sec.\,\ref{SecFaddeevE}; and detail the quark-quark interaction that we use in solving for all one-, two- and three-valence-body Schwinger functions relevant to our calculation of the proton's tensor charges in Sec.\,\ref{SeccalG}.  Sec.\,\ref{SecTensorCharge} introduces the proton's tensor current, explains how to extract the tensor charges therefrom, describes the gap equation for the dressed light-quark propagator and its solution, and presents the solution of the inhomogeneous Bethe-Salpeter equation for the dressed-quark-tensor vertex.  Our results for the proton's tensor charges are reported and discussed in Sec.\,\ref{protonTCNumerical}.  Sec.\,\ref{SecEDM} connects these results with nucleon EDMs and Sec.\,\ref{SecEpilogue} provides a summary and perspective.

%

\section{Three-body Amplitude and Equation}
\label{SecFaddeevE}
The Faddeev amplitude for the $J=1/2$ nucleon can be written as follows:
\begin{align}
\nonumber
 \, _{c_1 c_2 c_3} & \mathbf{\Psi}^{\alpha_1\alpha_2\alpha_3,\delta}_{\iota_1 \iota_2 \iota_3,\iota}(p_1,p_2,p_3;P)  \nonumber \\
& = \tfrac{1}{\surd 6} \varepsilon_{c_1 c_2 c_3}
{\Psi}^{\alpha_1\alpha_2\alpha_3,\delta}_{\iota_1 \iota_2 \iota_3,\iota}(p_1,p_2,p_3;P)  \,,
\label{FaddeevAmp}
\end{align}
where
$c_{1,2,3}$ are colour indices;
$\alpha_{1,2,3}$, $\delta$ are spinor indices for the three valence quarks and nucleon, respectively;
$\iota_{1,2,3}$, $\iota$ are analogous isospin indices;
and $P=p_1+p_2+p_3$, $P^2 = -M_{N}^2$, where $M_N$ is the nucleon mass and $p_{1,2,3}$ are the valence-quark momenta.

With colour factorised from the amplitude in Eq.\,\eqref{FaddeevAmp}, then
${\Psi}_{\iota_1 \iota_2 \iota_3,\iota}^{\alpha_1\alpha_2\alpha_3,\delta}(p_1,p_2,p_3;P)$ describes momentum-space$+$spin$+$isospin correlations in the nucleon and must be symmetric under the interchange of any two valence quarks, including cyclic permutations, \emph{e.g}.\
\begin{equation}
{\Psi}^{\alpha_1\alpha_2\alpha_3,\delta}_{\iota_1 \iota_2 \iota_3,\iota}(p_1,p_2,p_3;P) = {\Psi}^{\alpha_1\alpha_3\alpha_2,\delta}_{\iota_1 \iota_3 \iota_2,\iota}(p_1,p_3,p_2;P)\,,
\end{equation}
As we shall now describe, the structure of this matrix-valued function is nontrivial in a Poincar\'e-covariant treatment.

\begin{figure}[!t]
\begin{center}
\begin{tabular}{lcr}
\parbox[c]{0.35\linewidth}{\includegraphics[clip,width=\linewidth]{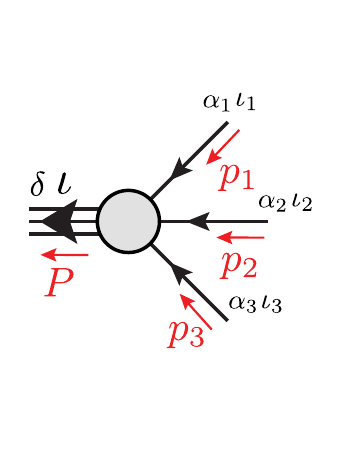}} &
 \hspace*{-1.5em} \mbox{\large $= \sum_{\{1,2,3\}}$} \hspace*{-1em} &
\parbox[c]{0.47\linewidth}{\includegraphics[clip,width=\linewidth]{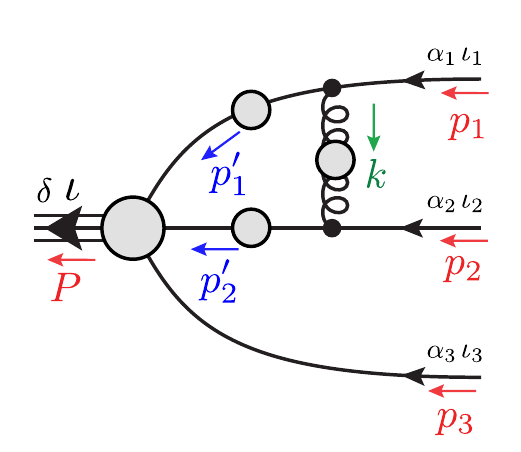}}\vspace*{-0ex}
\end{tabular}
\end{center}
\vspace*{-2em}
\caption{\label{FEimage}
Three-body equation in Eq.\,\eqref{eq:faddeev0}, solved herein for the proton's mass and bound-state amplitude.
Amplitude:  vertex on the left-hand-side;
spring with shaded circle: quark-quark interaction kernel in Eq.\,\eqref{KDinteraction};
and solid line with shaded circle: dressed-propagators for scattering quarks, obtained by solving a gap equation with the same interaction (Sec.\,\ref{GapEq}).
}
\end{figure}

Consider isospin first.  There are three valence quarks in the fundamental representation of $SU(2)$ and
\begin{equation}
2\otimes 2\otimes 2 = 4 \oplus 2 \oplus 2\,.
\end{equation}
The fully-symmetric $4$-dimensional irreducible representation (irrep) is associated with the $\Delta$-baryon and therefore ignored hereafter.  In terms of valence-quark flavours, the two mixed-symmetry $I=1/2$ $2$-dimensional irreps can be depicted thus:
\begin{equation}
\begin{array}{c|c|c}
       & I_z =\frac{1}{2} &  I_z = -\frac{1}{2} \\ \hline
 \mathsf{F}_0      & \tfrac{1}{\surd 2} (udu - duu) & \tfrac{1}{\surd 2} (udd - dud) \\
 \mathsf{F}_1      &  -\tfrac{1}{\surd 6}(udu+duu-2uud) & \tfrac{1}{\surd 6}(udd+dud-2ddu)
\end{array}\,.
\end{equation}
Defining a quark isospin vector $\mathsf{f}=(u,d)$, then this array can be expressed compactly via matrices:
\begin{equation}
\label{DiquarkSeeds}
D_0 = \tfrac{i}{\surd 2} \tau^2 \otimes \tau^0\,,\;
D_1 = -\tfrac{i}{\surd 6} \tau^i \tau^2 \otimes \tau^i\,,
\end{equation}
where $\tau^0 = {\rm diag}[1,1]$ and $\{\tau^i,i=1,2,3\}$ are Pauli matrices, \emph{e.g}.\ the bottom-left entry is
\begin{equation}
(\mathsf{f}\, \mathsf{f}^{\rm T}) D_1  (\mathsf{f}\, \mathsf{p}^{\rm T}) = -\tfrac{i}{\surd 6} \mathsf{f} \tau^i \tau^2 \mathsf{f}^{\rm T} \, \mathsf{f} \tau^i \mathsf{p}^{\rm T},
\end{equation}
where $\mathsf{p}=(1,0)$ represents the $I_z=+1/2$ proton.  Notably, with respect to the first two labels, $D_0$ relates to isospin-zero and $D_1$ to isospin-one; and differences between quark-quark scattering in these channels can provide the seed for formation of diquark correlations within baryons \cite{Segovia:2015ufa}.  Such differences do exist, \emph{e.g}.\ only $u$-$d$ scattering possesses an attractive isospin-zero channel.

Labelling the valence quarks by $\{i,j,k\}$, each taking a distinct value from $\{1,2,3\}$, then under $i \leftrightarrow j$
\begin{equation}
\left[\begin{array}{c}
\mathsf{F}_0 \\ \mathsf{F}_1 \end{array} \right]
\to
\left[\begin{array}{c}
\mathsf{F}_0^\prime \\ \mathsf{F}_1^\prime \end{array} \right]
=
{\mathpzc E}_k
\left[\begin{array}{c}
\mathsf{F}_0 \\ \mathsf{F}_1 \end{array} \right]\,,
\end{equation}
where ${\mathpzc E}_k$ is the associated exchange operator.  In general, owing to the mixed symmetry of these irreps, $\mathsf{F}_{0,1}^\prime \neq \mathsf{F}_{0,1}$.  Define in addition, therefore, a momentum-space$+$spinor doublet with the following transformation properties:
\begin{equation}
\left[ \Psi_0 \Psi_1 \right] \to \left[ \Psi_0 \Psi_1 \right] {\mathpzc E}_k^{\rm T}.
\end{equation}
The momentum-space$+$spinor$+$isospin combination
\begin{align}
\nonumber
&\Psi(p_1,p_2,p_3;P) \\
&= \Psi_0 (p_1,p_2,p_3;P) \mathsf{F}_0 + \Psi_1(p_1,p_2,p_3;P) \mathsf{F}_1
\label{FinalAmplitude}
\end{align}
is invariant under the exchange of any two quark labels.  (Given this ``doublet'' structure, $64+64=128$ independent scalar functions are required in general to completely describe a nucleon Faddeev amplitude: see Appendix\,B in Ref.\,\cite{Eichmann:2011vu} for more details.)  This feature is a statement of the fact that a Poincar\'e-covariant treatment of the nucleon does not typically admit a solution in which the momentum-space behaviour is independent of the spin-isospin structure; or, equivalently, that using a Poincar\'e-covariant framework, the $d$-quark contribution to a nucleon's form factor or kindred property is not simply proportional to the $u$-quark contribution.

Bound-states and their interactions can be studied in the continuum via a collection of coupled integral equations \cite{Roberts:1994dr}.  A tractable system of equations is only obtained once a truncation scheme is specified; and a systematic, symmetry-preserving approach is described in Refs.\,\cite{Munczek:1994zz, Bender:1996bb, Binosi:2016rxzd}.  The leading-order term is the rainbow-ladder (RL) truncation.  It is known to be accurate for ground-state light-quark vector- and isospin-nonzero-pseudoscalar-mesons, and related ground-state octet and decouplet baryons \cite{Chang:2011vu, Bashir:2012fs, Roberts:2015lja, Horn:2016rip, Eichmann:2016yit} because corrections largely cancel in these channels owing to the preservation of relevant Ward-Green-Takahashi identities ensured by the scheme \cite{Munczek:1994zz, Bender:1996bb, Binosi:2016rxzd}.
To obtain the nucleon amplitude in Eq.\,\eqref{FinalAmplitude}, we therefore consider the following RL-truncation three-body equation, depicted in Fig.\,\ref{FEimage}:
\begin{subequations}
 \label{eq:faddeev0}
\begin{align}
{\Psi}_{\iota_1 \iota_2 \iota_3,\iota}^{\alpha_1\alpha_2\alpha_3,\delta}&(p_1,p_2,p_3)
= \sum_{j=1,2,3} \big[ {\mathscr K} S S \Psi \big]_j\,,
\label{eq:faddeev01}\\
\big[ {\mathscr K} S S \Psi \big]_3 & =
\int_{dk} \mathscr{K}_{\iota_1 \iota_1^\prime \iota_2 \iota_2^\prime}^{\alpha_1\alpha_1',\alpha_2\alpha_2'}(p_1,p_2;p_1',p_2') \nonumber\\
& \rule{-4em}{0ex} \times
S_{\iota_1^\prime \iota_1^{\prime\prime}}^{\alpha_1'\alpha_2''}(p_1')
S_{\iota_2^\prime \iota_2^{\prime\prime}}^{\alpha_2'\alpha_2''}(p_2')
{\Psi}_{\iota_1'' \iota_2'' \iota_3;\iota}^{\alpha_1''\alpha_2''\alpha_3;\delta}(p_1',p_2',p_3) \,,
\end{align}
\end{subequations}
where $\int_{dk}$ represents a translationally-invariant definition of the four-dimensional integral and $\big[ {\mathscr K} S S \Psi\big]_{1,2}$ are obtained from $\big[ {\mathscr K} S S \Psi \big]_3$ by cyclic permutation of indices. 

\begin{figure}[!t]
\begin{eqnarray*}
\parbox{15mm}{\includegraphics[height=2cm]{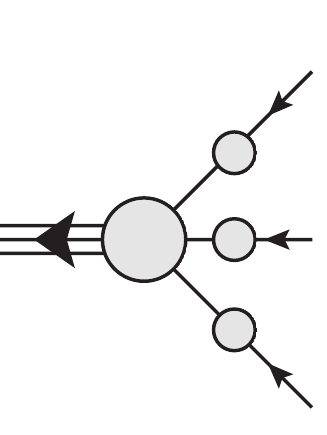}}
\left[\parbox{15mm}{\includegraphics[height=2cm]{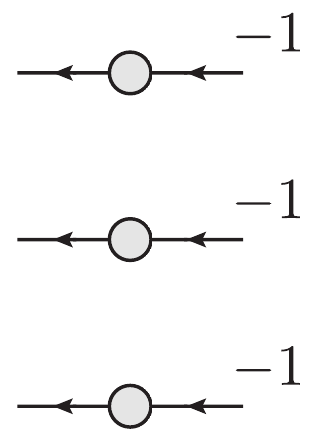}} \right]
\parbox{15mm}{\includegraphics[height=2cm]{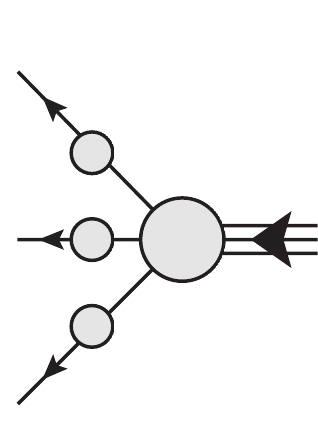}}
= \frac{1}{M} \left[\frac{\partial \lambda(P^2)}{\partial P^2} \right]^{-1}
\end{eqnarray*}
\caption{\label{canonicalnorm}
Evaluated on shell, \emph{i.e}.\ at $P^2=-M_N^2$, the bound-state amplitude which satisfies this identity is canonically normalised.
Amplitude:  vertices on either side of the square brackets; and
solid line with shaded circle: dressed-quark propagators (Sec.\,\ref{GapEq}).
The explicit appearance of the bracketed term with inverse propagators emphasises that the normalisation condition overlaps the amplitude with the unamputated bound-state wave function, either on the left or right.
}
\end{figure}

In order to compute any nucleon observable, the Faddeev amplitude must be canonically normalised.  This can be achieved by introducing an eigenvalue $\lambda(P^2) $ on the right-hand-side of Eq.\,\eqref{eq:faddeev01}.  The equation thus obtained has a solution at all values of $P^2$ and the original (bound-state) equation is recovered at that value of $P^2=-M_N^2$ for which $\lambda(-M_N^2) =1$.  Canonical normalisation is then achieved by rescaling the bound-state amplitude such that the identity in Fig.\,\ref{canonicalnorm} is satisfied at $P^2 = -M_N^2$ \cite{Nakanishi:1969ph}.  With such an amplitude, the proton has unit electric charge and the neutron is neutral.

\section{Two-Body Interaction}
\label{SeccalG}
The key element in analyses of the continuum bound-state problem for hadrons is the quark-quark scattering kernel.  In RL truncation that can be written ($k = p_1-p_1^\prime = p_2^\prime -p_2$):
\begin{subequations}
\label{KDinteraction}
\begin{align}
\mathscr{K}_{\alpha_1\alpha_1',\alpha_2\alpha_2'} & = {\mathpzc G}_{\mu\nu}(k) [i\gamma_\mu]_{\alpha_1\alpha_1'} [i\gamma_\nu]_{\alpha_2\alpha_2'}\,,\\
 {\mathpzc G}_{\mu\nu}(k) & = \tilde{\mathpzc G}(k^2) T_{\mu\nu}(k)\,,
\end{align}
\end{subequations}
where $k^2 T_{\mu\nu}(k) = k^2 \delta_{\mu\nu} - k_\mu k_\nu$.
Thus, in order to define all elements in Eq.\,\eqref{eq:faddeev0} and hence the bound-state problem, it remains only to specify $\tilde{\mathpzc G}$; and two decades of study have led to the following form \cite{Qin:2011dd, Qin:2011xq} ($s=k^2$):
\begin{align}
\label{defcalG}
 \tfrac{1}{Z_2^2}\tilde{\mathpzc G}(s) & =
 \frac{8\pi^2}{\omega^4} D e^{-s/\omega^2} + \frac{8\pi^2 \gamma_m \mathcal{F}(s)}{\ln\big[ \tau+(1+s/\Lambda_{\rm QCD}^2)^2 \big]}\,,
\end{align}
where: $\gamma_m=12/(33-2N_f)$, $N_f=4$; $\Lambda_{\rm QCD}=0.234\,$GeV; $\tau={\rm e}^2-1$; and ${\cal F}(s) = \{1 - \exp(-s/[4 m_t^2])\}/s$, $m_t=0.5\,$GeV.  $Z_2$ is the dressed-quark wave function renormalisation constant.
We employ a mass-independent momentum-subtraction renormalisation scheme for the gap and inhomogeneous vertex equations, implemented by making use of the scalar Ward-Green-Takahashi identity and fixing all renormalisation constants in the chiral limit \cite{Chang:2008ec}, with renormalisation scale $\zeta=2\,$GeV$=:\zeta_2$.  

The development of Eqs.\,\eqref{KDinteraction}, \eqref{defcalG} is summarised in Ref.\,\cite{Qin:2011dd} and their connection with QCD is described in Ref.\,\cite{Binosi:2014aea}; but it is worth reiterating some points..

The interaction in Eqs.\,\eqref{KDinteraction}, \eqref{defcalG} is deliberately consistent with that determined in studies of QCD's gauge sector, which indicate that the gluon propagator is a bounded, regular function of spacelike momenta that achieves its maximum value on this domain at $k^2=0$ \cite{Bowman:2004jm, Boucaud:2011ug, Ayala:2012pb, Aguilar:2012rz, Binosi:2014aea, Binosi:2016wcx, Binosi:2016xxu, Binosi:2016nme, Gao:2017uox}, and the dressed-quark-gluon vertex does not possess any structure which can qualitatively alter these features \cite{Skullerud:2003qu, Bhagwat:2004kj, Aguilar:2014lha, Bermudez:2017bpx}.  It also preserves the one-loop renormalisation group behaviour of QCD so that, \emph{e.g}.\ the quark mass-functions produced are independent of the renormalisation point.  On the other hand, in the infrared, \emph{i.e}.\ $k^2 \lesssim M_N^2$, Eq.\,\eqref{defcalG} defines a two-parameter model, the details of which determine whether confinement and/or dynamical chiral symmetry breaking (DCSB) are realised in solutions of the dressed-quark gap equations.

Computations \cite{Qin:2011dd, Qin:2011xq} reveal that observable properties of light-quark ground-state vector- and isospin-nonzero pseudoscalar-mesons are practically insensitive to variations of $\omega \in [0.4,0.6]\,$GeV, so long as
\begin{equation}
 \varsigma^3 := D\omega = {\rm constant}.
\label{Dwconstant}
\end{equation}
This feature also extends to numerous properties of the nucleon and $\Delta$-baryon \cite{Eichmann:2008ef, Eichmann:2012zz}.  The value of $\varsigma$ is chosen so as to obtain the measured value of the pion's leptonic decay constant, $f_\pi$; and in RL truncation this requires
\begin{equation}
\label{varsigmalight}
\varsigma  =0.80\,{\rm GeV.}
\end{equation}

It is also worth looking at Eq.\,\eqref{defcalG} from a different perspective \cite{Binosi:2014aea, Binosi:2016nme}.  Namely, one can sketch a connection with QCD's renormalisation-group-invariant process-independent effective charge by writing
\begin{equation}
\label{alphaRL}
\tfrac{1}{4\pi}\tilde{\mathpzc G}(s) \approx \frac{\tilde\alpha_{\rm PI}(s)}{s + \tilde m_g^2(s)}\,,\;
 m_g^2(s) = \frac{\tilde m_0^4}{s + \tilde m_0^2}\,,
\end{equation}
and extract $\tilde\alpha_{\rm PI}(0)=:\tilde\alpha_0$, $\tilde m_0$ via a least-squares fit on an infrared domain: $s\lesssim M_N^2$.  In this way, one obtains
\begin{equation}
\tfrac{1}{\pi}\tilde\alpha_0^{\rm RL} = 9.7\,,\; \tilde m_0^{\rm RL} = 0.54\,{\rm GeV}\,,\;
\end{equation}
$\alpha_0^{\rm RL}/\pi/[m_0^{\rm RL}]^2 \approx 33\,$GeV$^{-2}$.  Comparison of these values with those predicted via a combination of continuum and lattice analyses of QCD's gauge sector \cite{Binosi:2016nme}: $\alpha_0/\pi \approx 0.95$, $m_0 \approx 0.5\,$GeV, $\alpha_0/\pi/m_0^2 \approx 4.2\,$GeV$^{-2}$, confirms an earlier observation \cite{Binosi:2014aea} that the RL interaction defined by Eqs.\,\eqref{KDinteraction}, \eqref{defcalG} has the right shape, but is an order-of-magnitude too large in the infrared.  As explained elsewhere \cite{Chang:2009zb, Chang:2010hb, Chang:2011ei}, this is because Eq.\,\eqref{KDinteraction} suppresses all effects associated with DCSB in bound-state equations \emph{except} those expressed in $\tilde{\mathpzc G}(k^2)$, and therefore a description of hadronic phenomena can only be achieved by overmagnifying the gauge-sector interaction strength at infrared momenta.

In choosing the scale in Eq.\,\eqref{varsigmalight} so as to describe a given set of light-hadron observables in RL truncation, one also implicitly incorporates some of the effects of resonant corrections (meson cloud effects) on light-hadron static properties \cite{Eichmann:2008ae}.  We capitalise on this feature herein.

\begin{figure*}[!t]
\begin{eqnarray*}
\mbox{\large $[J^{(3)}_{nn'}(Q)]^{\delta\delta'}_{\mu\nu} = $}
\parbox{19em}{\includegraphics[width=\linewidth]{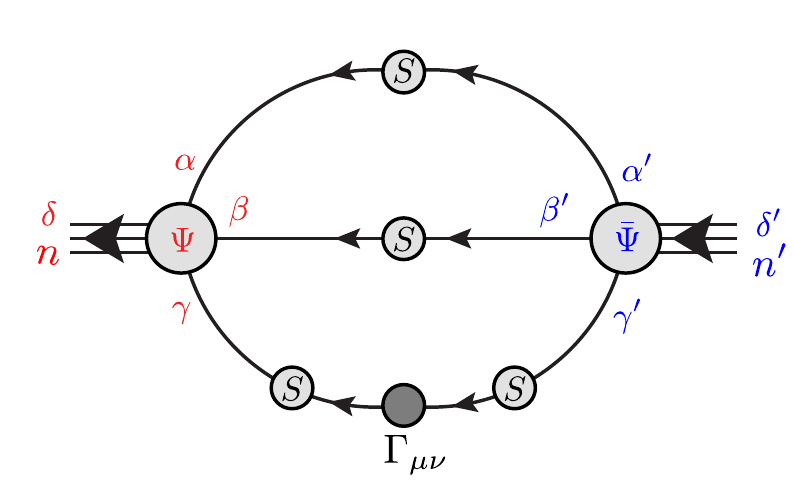}}
\mbox{\large $-$}
\parbox{19em}{\includegraphics[width=\linewidth]{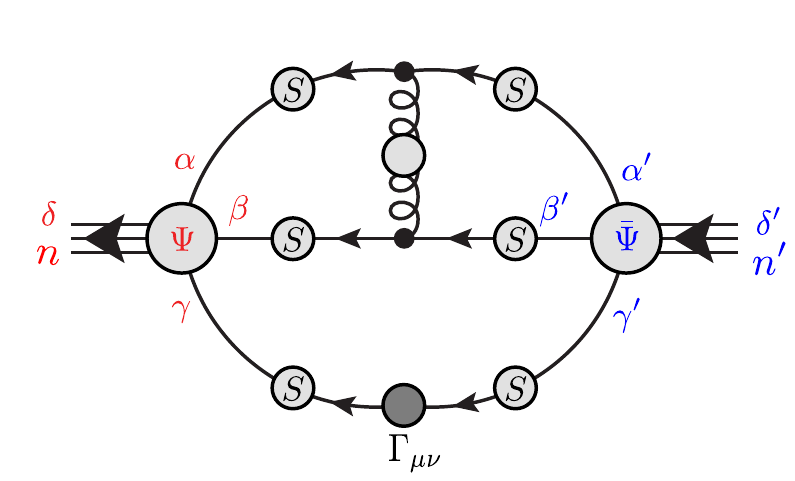}}\,.
\end{eqnarray*}
\vspace{-.75cm}	
\caption{\label{JmunuD}
$a=3$ spinor component of the tensor current in Eq.\,\eqref{tensorcurrent}: $\delta$, $\delta^\prime$ are spinor indices and $n$, $n^\prime$ are isospin indices: first term on the right-hand-side, impulse contribution; and second term, interaction correction.  The new element is the dressed-quark-tensor vertex, $\Gamma_{\mu\nu}$, described in Sec.\,\ref{SecTmunu}.
}
\end{figure*}

\section{Tensor Charges: Preliminaries}
\label{SecTensorCharge}
\subsection{Algebra}
Working with the definition of the tensor charge in Eq.\,\eqref{tcd} and a proton defined as a solution of Eq.\,\eqref{eq:faddeev0}, a symmetry-preserving calculation of the proton's tensor charge proceeds by computing the following current at zero momentum transfer:
\begin{align}
\label{tensorcurrent}
J_{\mu\nu}(Q) = \sum_{k=1}^{3}\sum_{nn'} [J^{(k)}_{nn'}(Q)]_{\mu\nu} \mathsf{F}^{(k)}_{nn'} \,,
\end{align}
where the spinor piece is illustrated in Fig.\,\ref{JmunuD} and the quantities $\{\mathsf{F}^{(k)}_{nn'}, k=1,2,3\}$ express the correlated isospin traces, \emph{e.g}.\
\begin{align}
	\mathsf{F}^{(3)}_{nn'}(\mathsf{p}, \mathsf{f}) = [D_{n'}^\dagger]_{bad'c'} \mathsf{f}_{c'}\mathsf{p}_{d'} \, [D_{n}]_{abcd}\mathsf{f}_{c} \mathsf{p}_{d}
\end{align}
isolates the contribution from flavour $\mathsf{f}$ to the proton's tensor charge: $\mathsf{u}=(1,0)$, $\mathsf{d}=(0,1)$.
%
Completing the algebra,
\begin{subequations}
\begin{align}
\mathsf{F}^{(3)}_{nn'}(\mathsf{p}, \mathsf{u}) & =
\left[ \begin{array}{cc}
1 & 0 \\
0 & \frac{1}{3}
\end{array}
\right]_{n n^\prime},\\
\mathsf{F}^{(3)}_{nn'}(\mathsf{p}, \mathsf{d}) & =
\left[ \begin{array}{cc}
0 & 0 \\
0 & \frac{2}{3}
\end{array}
\right]_{n n^\prime}\,.
\end{align}
\end{subequations}
and
\begin{subequations}
\begin{align}
\delta^{3}_T u & = \mathsf{J}^{(3)}_{00} + \tfrac{1}{3} \mathsf{J}^{(3)}_{11} \,,\\
\delta^{3}_T d & = \tfrac{2}{3} \mathsf{J}_{11}^{(3)} \,,
\end{align}
\end{subequations}
where, with the trace over spinor indices,
\begin{equation}
\label{Jnn3}
\mathsf{J}_{n n^\prime}^{(3)} = {\rm tr} \tfrac{1}{12} \sigma_{\mu\nu}  [J^{(3)}_{nn'}(0)]_{\mu\nu}\,.
\end{equation}
Now, owing to symmetry of the colour-factorised proton wave function under interchange of any two quarks, one arrives at the final result:
\begin{subequations}
\label{udTensorCharges}
\begin{align}
\delta_T u & = 3 \delta^{3}_T u = 3 \mathsf{J}^{(3)}_{00} + \mathsf{J}^{(3)}_{11} \,,\\
\delta_T d & = 3 \delta^{3}_T d = 2 \mathsf{J}_{11}^{(3)} \,.
\end{align}
\end{subequations}
Recalling the conclusions of Refs.\,\cite{Pitschmann:2014jxa, Xu:2015kta} and the remarks following Eq.\,\eqref{DiquarkSeeds}, it is evident from this analysis that $\delta_T d \simeq 0$ in any model of nucleon structure that retains only scalar diquark correlations.

The algebraic structure in Eqs.\,\eqref{udTensorCharges} is quite general.  For instance, considering the nucleon axial charges in the isospin symmetric limit:
\begin{subequations}
\begin{align}
A_{uu}^{pp} & = \langle P(k,\sigma) | \bar u \gamma_5\gamma_\mu u |  P(k,\sigma) \rangle\,,\\
A_{dd}^{pp} & = \langle P(k,\sigma) | \bar d \gamma_5\gamma_\mu d |  P(k,\sigma) \rangle\,,\\
A_{ud}^{pn} & = \langle P(k,\sigma) | \bar u \gamma_5\gamma_\mu d|  N(k,\sigma) \rangle\,,
\end{align}
\end{subequations}
with $|N(k,\sigma)\rangle$ a neutron state vector, then
\begin{subequations}
\begin{align}
A_{uu}^{pp} & = 3 \mathsf{A}^{(3)}_{00} + \mathsf{A}^{(3)}_{11} \,,\\
A_{dd}^{pp} & = 2 \mathsf{A}_{11}^{(3)}\,,  \\
A_{ud}^{pn} & = A_{uu}^{pp} - A_{dd}^{pp}\,,
\end{align}
\end{subequations}
where
\begin{equation}
\mathsf{A}^{(3)}_{nn^\prime} = {\rm tr}\tfrac{1}{4} \gamma_\mu\gamma_5  [J^{(3)}_{nn'}(0)]_{5\mu}\,,
\end{equation}
with $[J^{(3)}_{nn'}(0)]_{5\mu}$ obtained from the current in Fig.\,\ref{JmunuD} by making the replacement $\Gamma_{\mu\nu} \to \Gamma_{5\mu}$, the latter being the dressed-quark-axial-vector vertex \cite{Maris:1997hd, GutierrezGuerrero:2010md, Chang:2012cc, Yamanaka:2014lva, Eichmann:2011pv}.

\subsection{Dressed-quark Propagator}
\label{GapEq}
The kernel of Eq.\,\eqref{eq:faddeev0}, the Faddeev equation, is complete once the dressed-quark propagator is known.  In order to ensure a symmetry-preserving analysis, this should be computed from the following (rainbow-truncation) gap equation:
\begin{subequations}
\label{gendseN}
\begin{align}
S^{-1}(k) & = i\gamma\cdot k \, A(k^2) + B(k^2) \\
%
& = Z_2 \,(i\gamma\cdot k + m^{\rm bm}) + \Sigma(k)\,,\\
\Sigma(k)& = \int_{dq}\!\!
 {\mathpzc G}_{\mu\nu}(k-q)\frac{\lambda^a}{2}\gamma_\mu S(q) \frac{\lambda^a}{2} \gamma_\nu \,,
\end{align}
\end{subequations}
using the interaction specified in connection with Eqs.\,\eqref{KDinteraction}, \eqref{defcalG}.  Following Ref.\,\cite{Maris:1997tm}, this gap equation is now readily solved, and we adapt algorithms from Ref.\,\cite{Krassnigg:2009gd} when necessary.  Solving the gap equation subject to the condition that the mass function reproduce  renormalisation-group-invariant current-quark masses
\begin{equation}
\hat m_u = \hat m_d = 6.6\,{\rm MeV},
\end{equation}
which correspond to one-loop evolved values $m_u^{\zeta_2} = m_d^{\zeta_2} = 4.4\,$MeV, a good description of $\pi$- and $\rho$-meson properties is obtained; and we use these values herein.

\begin{figure}[t!]
\centering
\includegraphics[width=0.92\linewidth]{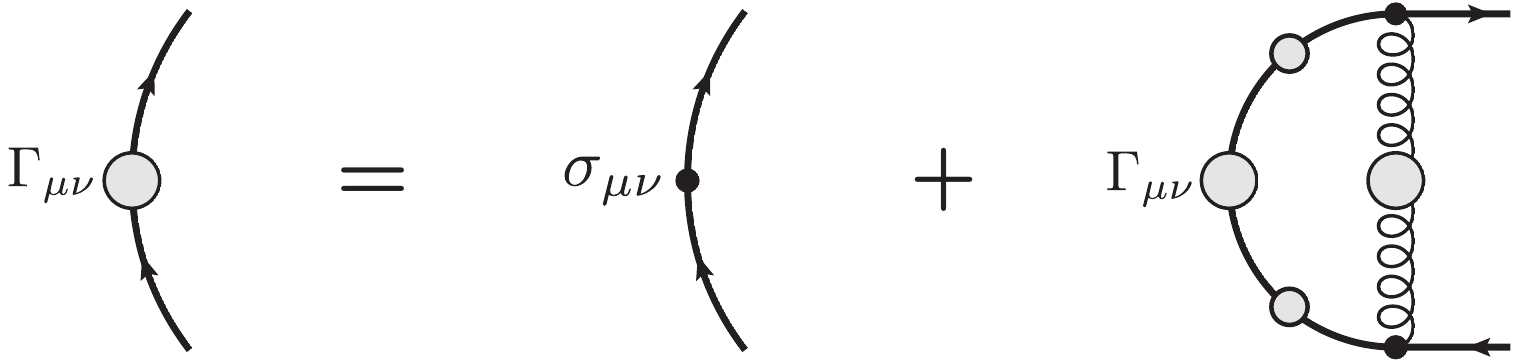}
\caption{\label{Figtsbse}
Inhomogeneous Bethe-Salpeter equation for the dres\-sed-quark-tensor vertex, $\Gamma_{\mu\nu}$, in rainbow-ladder truncation, which uses $S(k)$ from Eq.\,\eqref{gendseN}.}
\end{figure}

\subsection{Dressed-Quark-Tensor Vertex}
\label{SecTmunu}
The remaining element required to complete a calculation of the proton's tensor charges is the dressed-quark-tensor vertex, which satisfies the following inhomogeneous integral equation, depicted in Fig.\,\ref{Figtsbse}:
\begin{align}
\nonumber
&\Gamma_{\mu\nu}(k;Q)  = Z_T \sigma_{\mu\nu}  \\
&+ \int_{dq}\!\!
{\mathpzc G}_{\mu\nu}(k-q)\frac{\lambda^a}{2}\gamma_\mu S(q_+) \Gamma_{\mu\nu}(q;Q) S(q_-)\frac{\lambda^a}{2} \gamma_\nu \,,
\label{IHBSET}
\end{align}
where $q_\pm = q\pm Q/2$ and $Z_T$ is the tensor vertex renormalisation constant, ensuring $\Gamma_{\mu\nu}(k^2=\zeta_2^2;Q=0) = \sigma_{\mu\nu}$.

\begin{figure}[t!]
\centering
\includegraphics[width=0.87\linewidth]{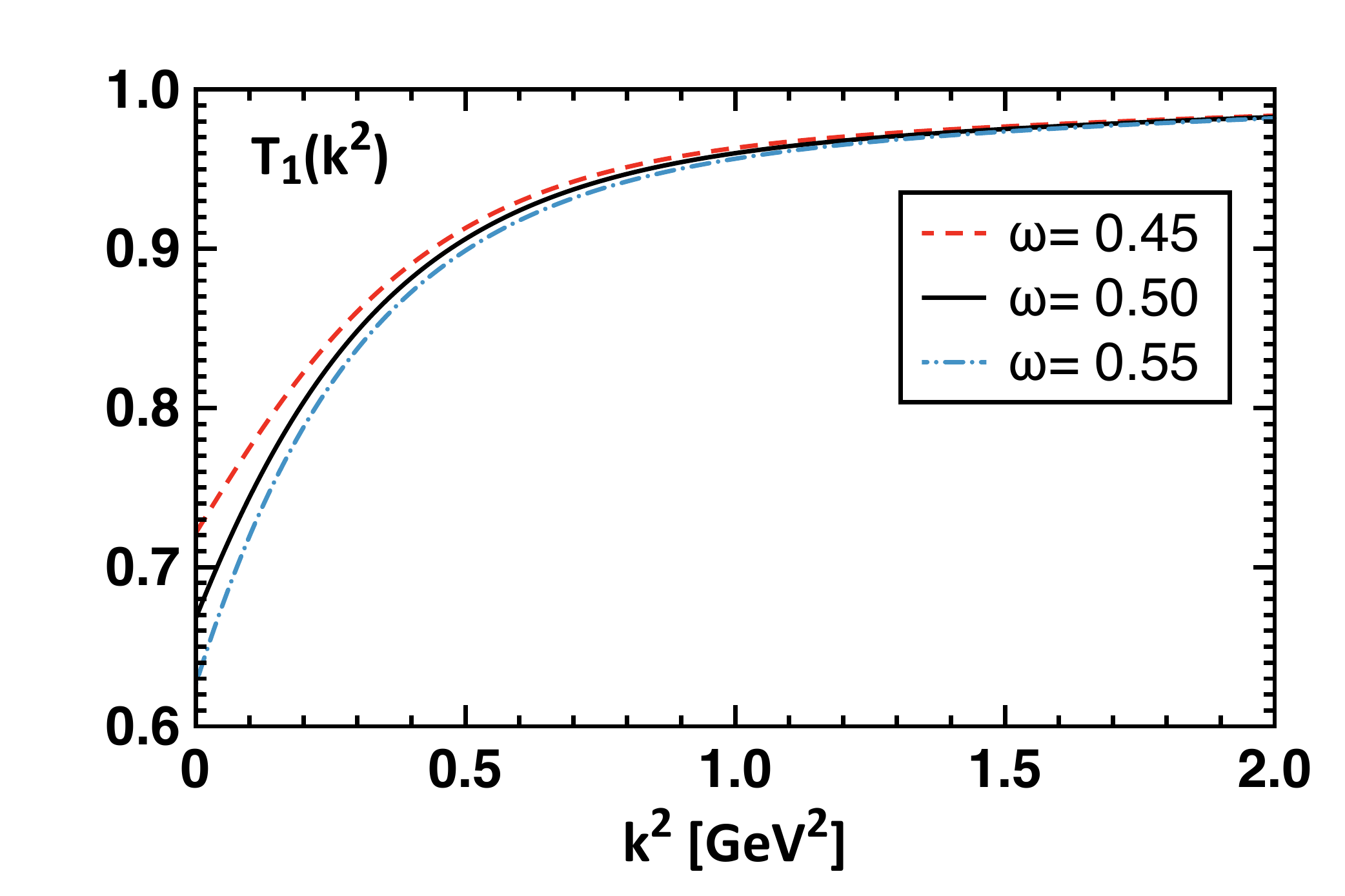}
\includegraphics[width=0.87\linewidth]{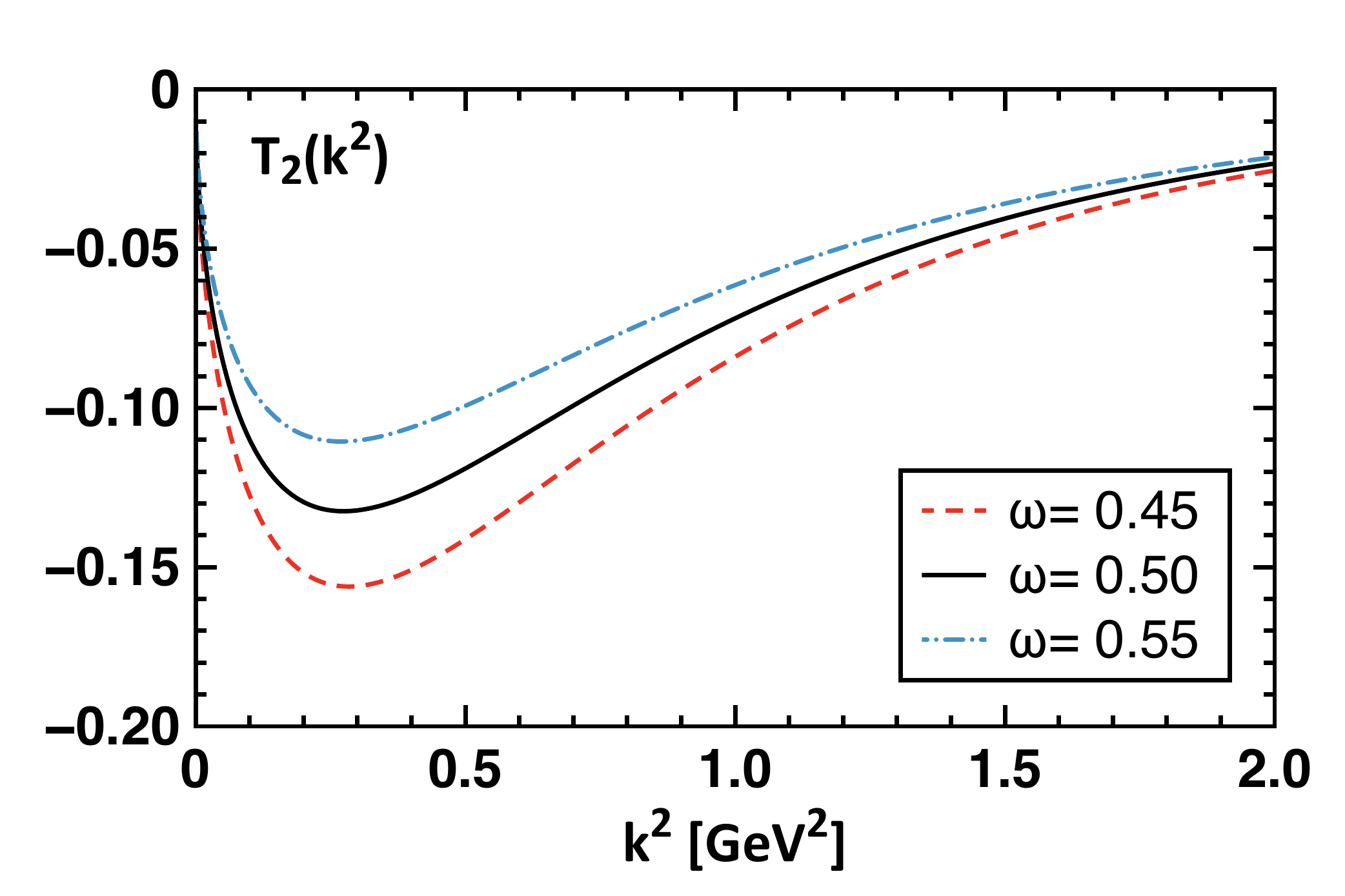}
\includegraphics[width=0.87\linewidth]{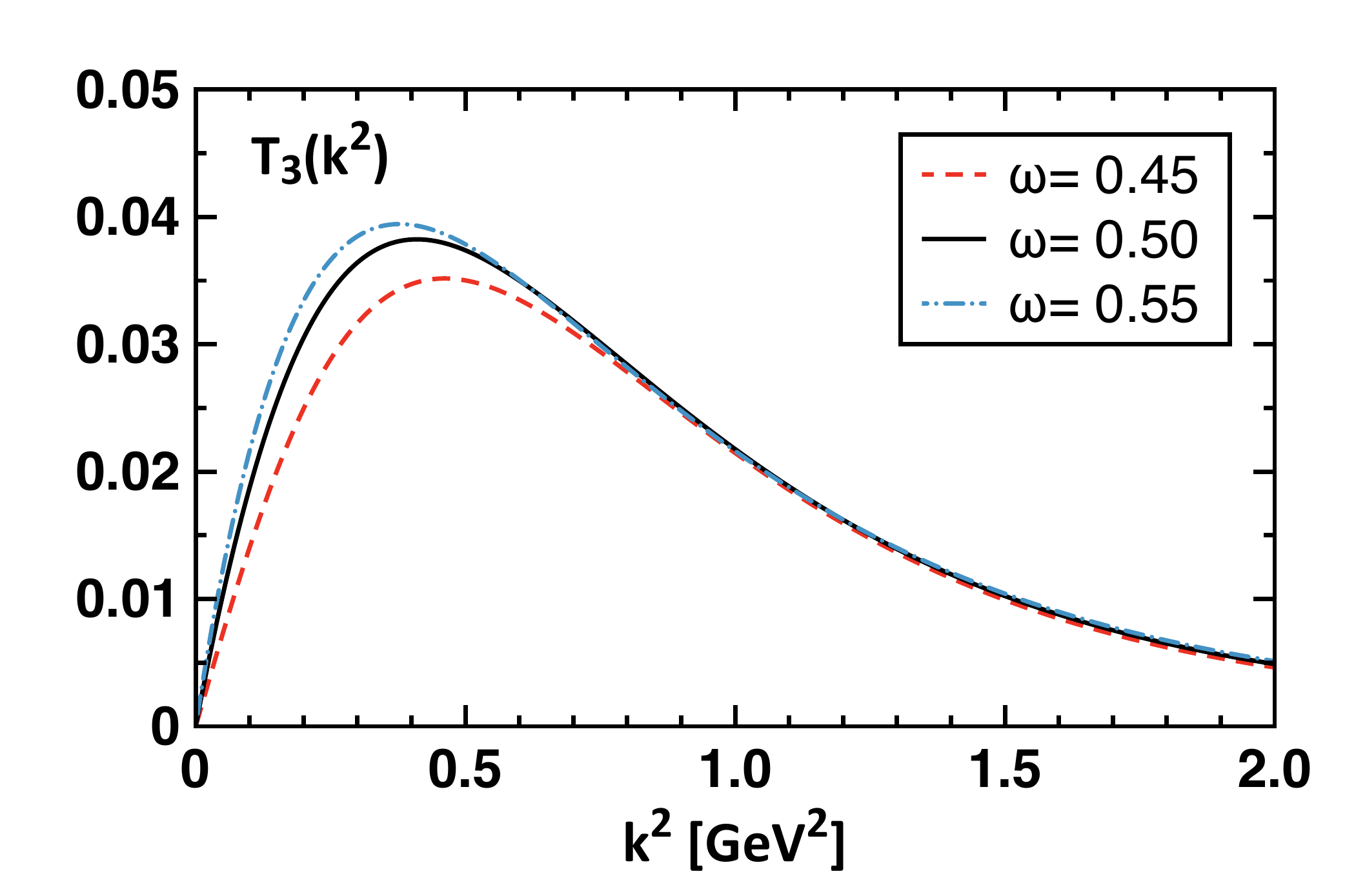}
\caption{\label{S123}
Scalar functions determining the $Q=0$ dressed-quark-tensor vertex in Eq.\,\eqref{Gammamunu0} obtained using the interaction in Eq.\,\eqref{defcalG} and $S(p)$ computed from Eq.\,\eqref{gendseN}.  Shown, too, is the sensitivity to a $\pm 10$\% variation in the interaction's range parameter, subject to Eqs.\,\eqref{Dwconstant}, \eqref{varsigmalight}.}
\end{figure}

Computation of the proton's tensor charge only requires knowledge of
\begin{align}
\nonumber
\Gamma_{\mu\nu}(k;Q=0) & = T_1(k^2;\zeta) \sigma_{\mu\nu}
+ T_2(k^2;\zeta) \{ \gamma\cdot \hat k , \sigma_{\mu\nu} \} \\
&
+ T_3(k^2;\zeta) (\sigma_{\mu\rho} \hat k_\rho \hat k_\nu - \sigma_{\nu\rho}\hat k_\rho \hat k_\mu )\,,
\label{Gammamunu0}
\end{align}
where $\hat p^2 = 1$.  Inserting Eq.\,\eqref{Gammamunu0} into Eq.\,\eqref{IHBSET}, one obtains a set of three coupled linear integral equations for $T_{1,2,3}$, whose kernels are completely specified by $\tilde{\mathpzc G}$ in Eq.\,\eqref{defcalG} and $S(k)$ computed using Eq.\,\eqref{gendseN}.  The solutions are depicted in Figs.\,\ref{S123} and are characterised by the following value of the light-quark tensor charge:
\begin{equation}
\label{quarkT}
T_1(k^2=0;\zeta_2) = 0.67(5) =: \tilde \delta_T q\,.
\end{equation}
This value may be compared with the estimate reported in Ref.\,\cite{Yamanaka:2013zoa}: $T_1(0;\zeta_2) \approx 0.6$.

It is worth remarking that DCSB leads similarly to a suppression of a dressed-quark's axial charge, but the effect is weaker, \emph{viz}.\ in comparison with the undressed value of unity, $\tilde g_A^q \approx 0.85$ \cite{Maris:1997hd, GutierrezGuerrero:2010md, Chang:2012cc, Yamanaka:2014lva}.  The missing strength is absorbed by the pion bound-state \cite{Maris:1997hd, GutierrezGuerrero:2010md, Chang:2012cc}.  The difference between $\tilde \delta_T q$ and $\tilde g_A^q$ highlights again the importance of preserving Poincar\'e-covariance in treatments of light-quark bound-state problems.

\section{Tensor Charges: Results and Analysis}
\label{protonTCNumerical}
Everything necessary to evaluate the proton's tensor charges is now available.
To proceed, we solve the three-body equation, Eq.\,\eqref{eq:faddeev0}, for the proton's mass and bound-state amplitude, using the interaction described in Sec.\,\ref{SeccalG} and the dressed-quark propagator from Sec.\,\ref{GapEq}, with the result ($\omega = 0.5 \mp 0.05$)
\begin{equation}
m_{N} \,({\rm GeV}) = 0.932^{(5)}_{(11)}\,.
\end{equation}
Importantly, no parameters were varied to obtain this value: it follows once the scale in Eq.\,\eqref{Dwconstant} is chosen.  Using this amplitude, canonically normalised as described in connection with Fig.\,\ref{canonicalnorm}, along with the same interaction and quark propagator, and the dressed-quark-tensor vertex described in Sec.\,\ref{SecTmunu}, we compute $\mathsf{J}^{(3)}$ in Eq.\,\eqref{Jnn3} from the current in Fig.\,\ref{JmunuD}.  Subsequently, using Eqs.\,\eqref{udTensorCharges} $(\omega = 0.5 \mp 0.05)$:
\begin{subequations}
\label{charges}
\begin{align}
\label{deltaTproton}
& \delta_T u  = 0.912^{(42)}_{(47)} \,,\; & \delta_T d & = - 0.218^{(4)}_{(5)}\,,\\
\label{gT1proton}
& g_T^{(1)} = 1.130^{(42)}_{(47)} \,,\; &  g_T^{(0)} & = \phantom{-} 0.694^{(42)}_{(47)} \,.
\end{align}
\end{subequations}

It is interesting to note that if the dressed-quark tensor charge from Eq.\,\eqref{quarkT} is used in combination with a simple quark-model spin-flavour wave function \cite{He:1994gz, Yamanaka:2013zoa}, then one finds:
\begin{subequations}
\begin{align}
\delta_T^{\rm QM} u & = \phantom{-}\frac{4}{3} \tilde \delta_T q = \phantom{-} 0.89\,,\\
\delta_T^{\rm QM} d & =-\frac{1}{3} \tilde \delta_T q = -0.22 \,,
\end{align}
\end{subequations}
values which are practically equivalent to those in Eq.\,\eqref{deltaTproton}.  This similarity is a numerical accident, however.  If one instead uses the bare tensor vertex, so that the computed charges are a direct measure of proton wave function properties, then $\delta_T u = 1.12$, $ \delta_T d = -0.25$.

It is here worth noting that the quark model itself predicts \cite{He:1994gz} $ \delta_T^{\rm QM} u = (4/3)  = 1.33 $, $ \delta_T^{\rm QM} d = (-1/3)  = -0.33 $, values which should be associated with the ``model scale'', $\zeta_M$.  The model does not have a traceable connection with QCD so this scale is unknown; but $\zeta_M$ can be introduced as a parameter and tuned in order to obtain a desired result for the tensor charges.  On physical grounds, one should require $\zeta_M \gtrsim 2 \Lambda_{\rm QCD}$ so that perturbative evolution is possibly applicable \cite{Holt:2010vj}.  As noted in Ref.\,\cite{Xu:2015kta}, with $\zeta_M = 0.39\,$GeV one obtains $\delta_T^{\rm QM} u = 1.06$, $\delta_T^{\rm QM} d = -0.26$ using first-order evolution to reach $\zeta_2$.  A smaller value of $\zeta_M$ is difficult to justify.

\begin{figure}[t!]
\centering
\includegraphics[width=0.95\linewidth]{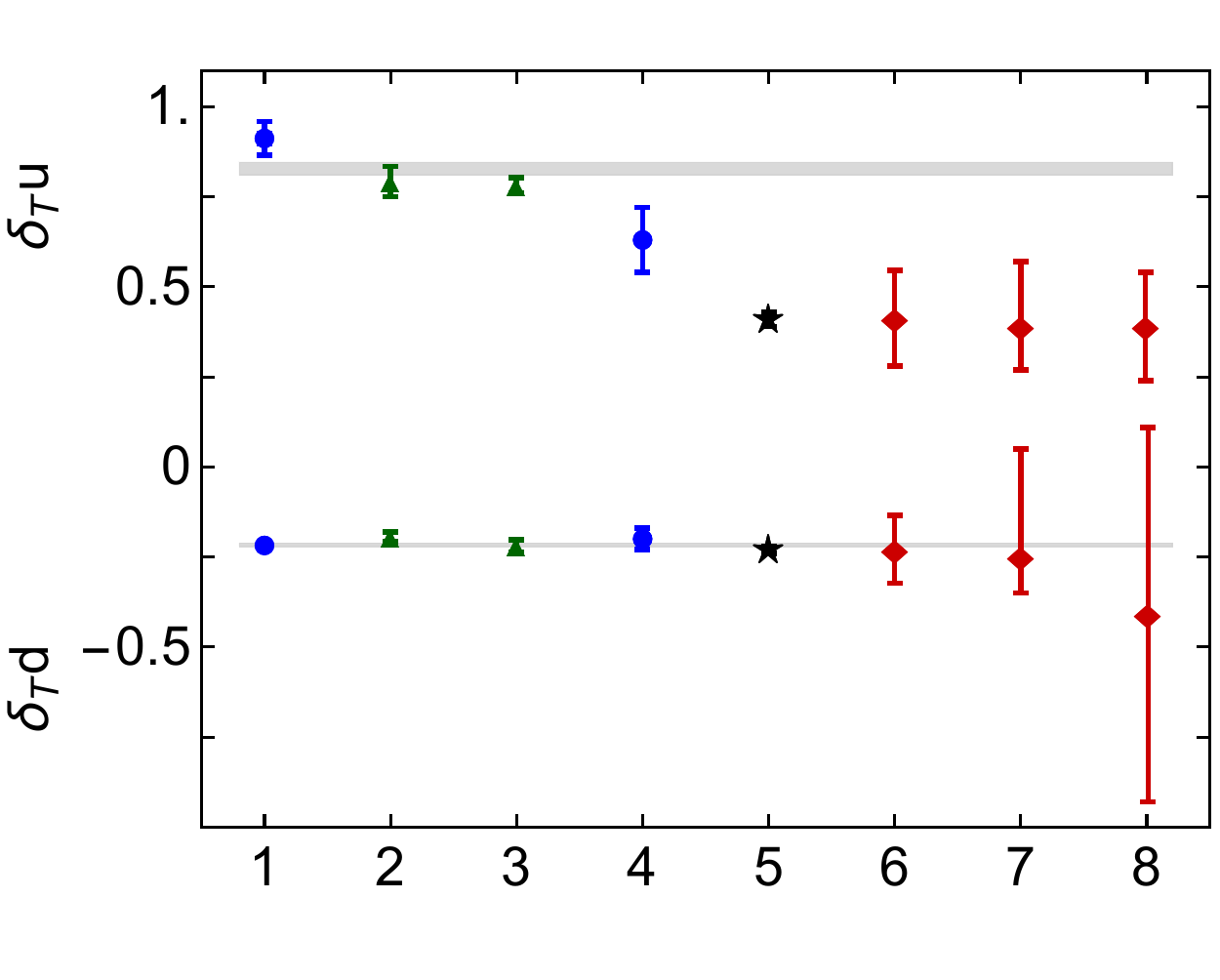}
\caption{\label{TCcompare}
Comparison of our prediction for the proton's tensor charges, position 1 -- Eq\,\eqref{deltaTproton}, with those obtained using:
lQCD (2 -- \cite{Bhattacharya:2016zcn} and 3 -- \cite{Alexandrou:2017qyt});
and a contact-interaction Faddeev equation (4 -- \cite{Xu:2015kta}).  The renormalisation scale is $\zeta^2 = 4\,$GeV$^2$ in all these cases; and the grey bands depict the averages in Eq.\,\eqref{DSElQCD}.
Position 5 -- projected errors achievable at JLab\,12 with the Solenoidal Large Intensity Device (SoLID) \cite{Gao:2017ade}, using Eq.\,\eqref{DefineTensorCharge} and anticipated transversity distribution data.  The central values are chosen to match those estimated elsewhere \cite{Ye:2016prn} (6, $\zeta^2=2.4\,$GeV$^2$) following an analysis of extant transversity distribution data.
Earlier estimates from transversity distribution data are also depicted (7 -- \cite{Anselmino:2013vqa}, $\zeta^2=2.4\,$GeV$^2$, and 8 -- \cite{Radici:2015mwa}, $\zeta^2=1\,$GeV$^2$.)
}
\end{figure}

In Fig.\,\ref{TCcompare} we compare our predictions for the proton's tensor charges with those obtained using lQCD \cite{Bhattacharya:2016zcn, Alexandrou:2017qyt} and an earlier contact-interaction Faddeev equation study \cite{Xu:2015kta}.  A weighted combination of our result and the most recent lQCD values \cite{Bhattacharya:2016zcn, Alexandrou:2017qyt} yields the following estimates (grey bands in Fig.\,\ref{TCcompare}):
\begin{equation}
\label{DSElQCD}
\overline{\delta_T} u  = 0.803(17)\,,\;  \overline{\delta_T} d  = - 0.216(4)\,.\\
\end{equation}
It is evident from the figure that our predictions are consistent with recent lQCD results; but these three analyses, based on Eq.\,\eqref{tcd}, produce results for $\delta_T u$ which differ markedly from those obtained via Eq.\,\eqref{DefineTensorCharge} using extant transversity distribution data.  (Tensor charges decrease with increasing $\zeta$, hence evolving the phenomenological estimates so that their renormalisation scales match the calculations would increase the discrepancy.)

It is here worth describing the fundamental differences between this study and that in Ref.\,\cite{Xu:2015kta}.
We use a single interaction kernel, Eq.\,\eqref{defcalG}, whose sole parameter is fixed by requiring a good description of $\pi$- and $\rho$-meson properties, to compute every element that contributes to the proton's tensor charge: the dressed-quark propagator; dressed-quark-tensor vertex; and proton Faddeev amplitude and tensor current.  On the domain of momenta for which perturbative-QCD is a valid tool, the behaviour of each of these quantities matches that required by QCD.  Their properties at infrared momenta are a direct reflection of the kernel's extension to that domain.  This is where the model-parametrisation has an influence; but progress is being made toward eliminating that as more is learnt about the infrared behaviour of Schwinger functions in strong QCD and how that information may be incorporated into the continuum bound-state problem \cite{Chang:2009zb, Chang:2010hb, Chang:2011ei, Binosi:2014aea, Williams:2015cvx, Binosi:2016wcx, Binosi:2016xxu, Binosi:2016nme, Gao:2017uox}.  Diquark correlations play no explicit role in our analysis.

On the other hand, Ref.\,\cite{Xu:2015kta} uses a symmetry-preserving (rainbow-ladder-like) treatment of a vector-vector contact interaction to develop and solve a Faddeev equation for the proton, in which isoscalar-scalar and isovector-vector diquark correlations play a key role and dynamical quark exchange between the diquarks provides an important contribution to binding within the nucleon.
The results obtained depend upon the values of seven parameters and the choice of regularisation scheme, the latter because a contact interaction is not renormalisable; and the unrealistic hardness of the interaction is expressed in many aspects of the results.  Notwithstanding these weaknesses, the algebraic simplicity of the framework enables some important qualitative features of many low-momentum observables to be clearly exhibited, such as the roles played by both DCSB and correlations in wave functions.  Naturally, where quantitative disagreements are met, one should prefer the QCD-connected results herein.


As noted in closing Sec.\,\ref{SeccalG}, the effects of improvements to RL truncation on some static hadron properties are implicitly included in the choice of scale, Eq.\,\eqref{varsigmalight}.  The residual dependence on $\omega$-variations can then be used to indicate just which static properties these might be \cite{Qin:2011dd}.  $M_{N}$ exhibits a 1\% response to variations $\omega \to (\omega \pm \Delta\omega)$, $\Delta\omega/\omega=0.1$, validating our approach to this observable.
$\delta_T u$ displays a $\lesssim 5$\% response to the same variation, flagging this as a quantity that might be sensitive to RL corrections.  In this case, its proximity to the independently obtained lQCD results may be used to argue for its stability; but we choose to look more deeply.
$\delta_T d$ is less sensitive to $\Delta\omega$, changing by only 2\%, which indicates that $\mathsf{J}_{11}^{(3)}$ in Eq.\,\eqref{udTensorCharges} is stable and hence the variation in $\delta_T u$ owes largely to that of $\mathsf{J}_{00}^{(3)}$.
An analogous feature is seen in Refs.\,\cite{Pitschmann:2014jxa, Xu:2015kta}, which employ a diquark approximation to the quark-quark scattering kernel and whose simplicity enables them to correlate the magnitude of $\delta_T u$ with the strength of DCSB as expressed, \emph{e.g}.\ in the nucleon mass and the integrated strength of the dressed-quark mass function.  In improving upon RL truncation, these things do not change, only the distribution and size of terms in the kernels which contribute to them \cite{Chang:2010hb, Chang:2011ei, Binosi:2014aea}.  Thus informed, one can more confidently rely upon the parameter-dependence results we have supplied as a valid estimate of the uncertainty in our tensor charge predictions.

\begin{figure}[t!]
\centering
\includegraphics[width=0.95\linewidth]{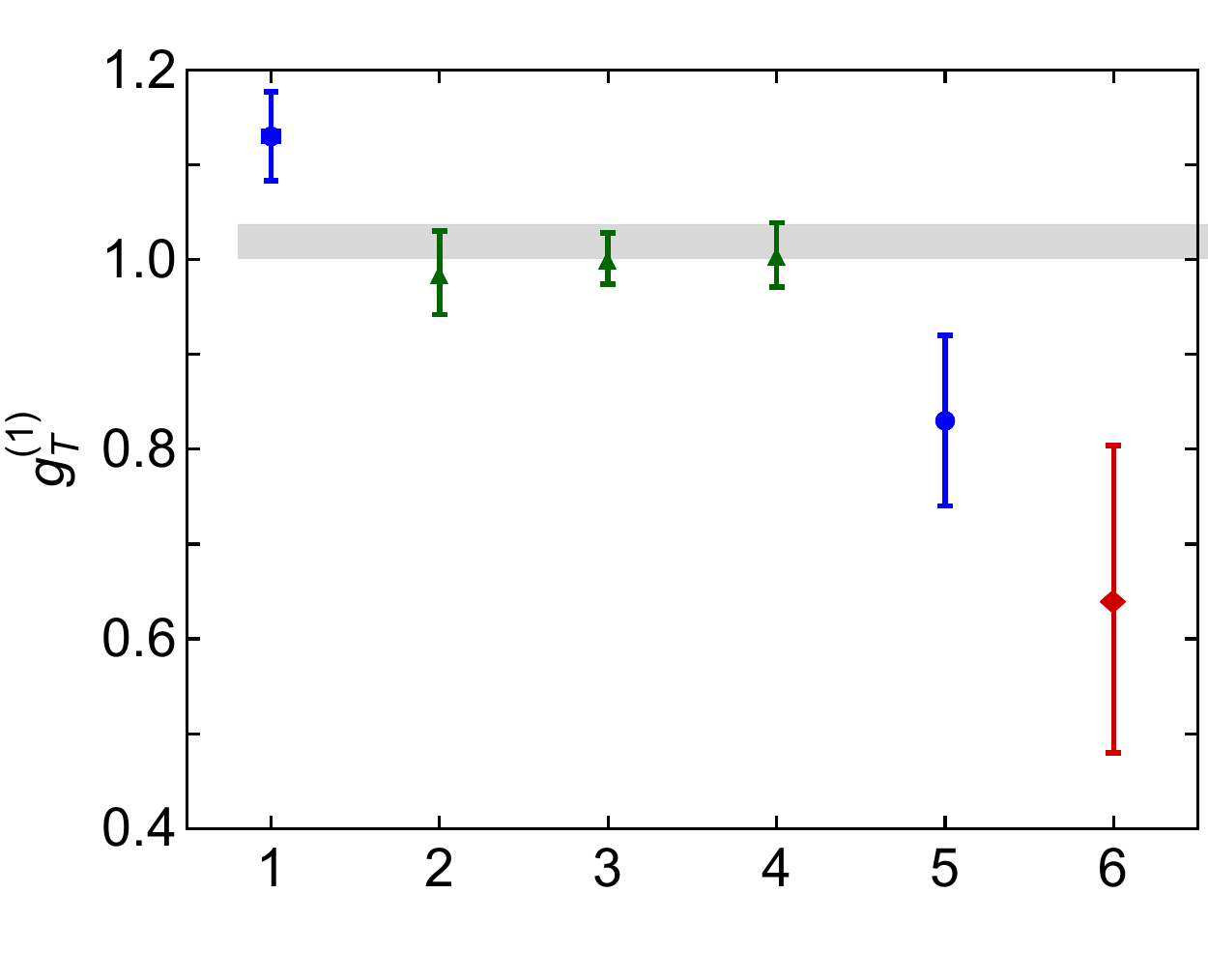}
\caption{\label{gTcompare}
Comparison of our prediction for the proton's isovector tensor charge (position 1 -- Eq\,\eqref{gT1proton}) with those obtained using:
lQCD (2 -- \cite{Bhattacharya:2016zcn},  3 -- \cite{Alexandrou:2017qyt}, 4 -- \cite{Bali:2014nma});
and a contact-interaction Faddeev equation (5 -- \cite{Xu:2015kta}).  The renormalisation scale is $\zeta^2 = 4\,$GeV$^2$ in all these cases; and the grey band depicts the weighted average in Eq.\,\eqref{DSElQCDgT}.
Position 6 -- estimate obtained using Eq.\,\eqref{DefineTensorCharge} and extant transversity distribution data \cite{Ye:2016prn} ($\zeta^2=2.4\,$GeV$^2$) .
}
\end{figure}

In Fig.\,\ref{gTcompare}, we compare our prediction for $g_T^{(1)}$ with values obtained by other means.  A weighted combination of our result and recent lQCD values \cite{Bhattacharya:2016zcn, Alexandrou:2017qyt, Bali:2014nma} yields the following estimate (grey bands in Fig.\,\ref{gTcompare}):
\begin{equation}
\label{DSElQCDgT}
\overline{g}_T^{(1)}  = 1.020(18)\,.
\end{equation}
The mismatch between theory and phenomenology is also apparent in this isovector combination of flavour-separated tensor charges.
(Recall that were a non-relativistic limit valid, then $g_T^{(1)}$ would match the nucleon's axial charge $g_A = 1.276$ \cite{Mendenhall:2012tz, Mund:2012fq}.)

\begin{figure}[t!]
\centering
\includegraphics[width=0.95\linewidth]{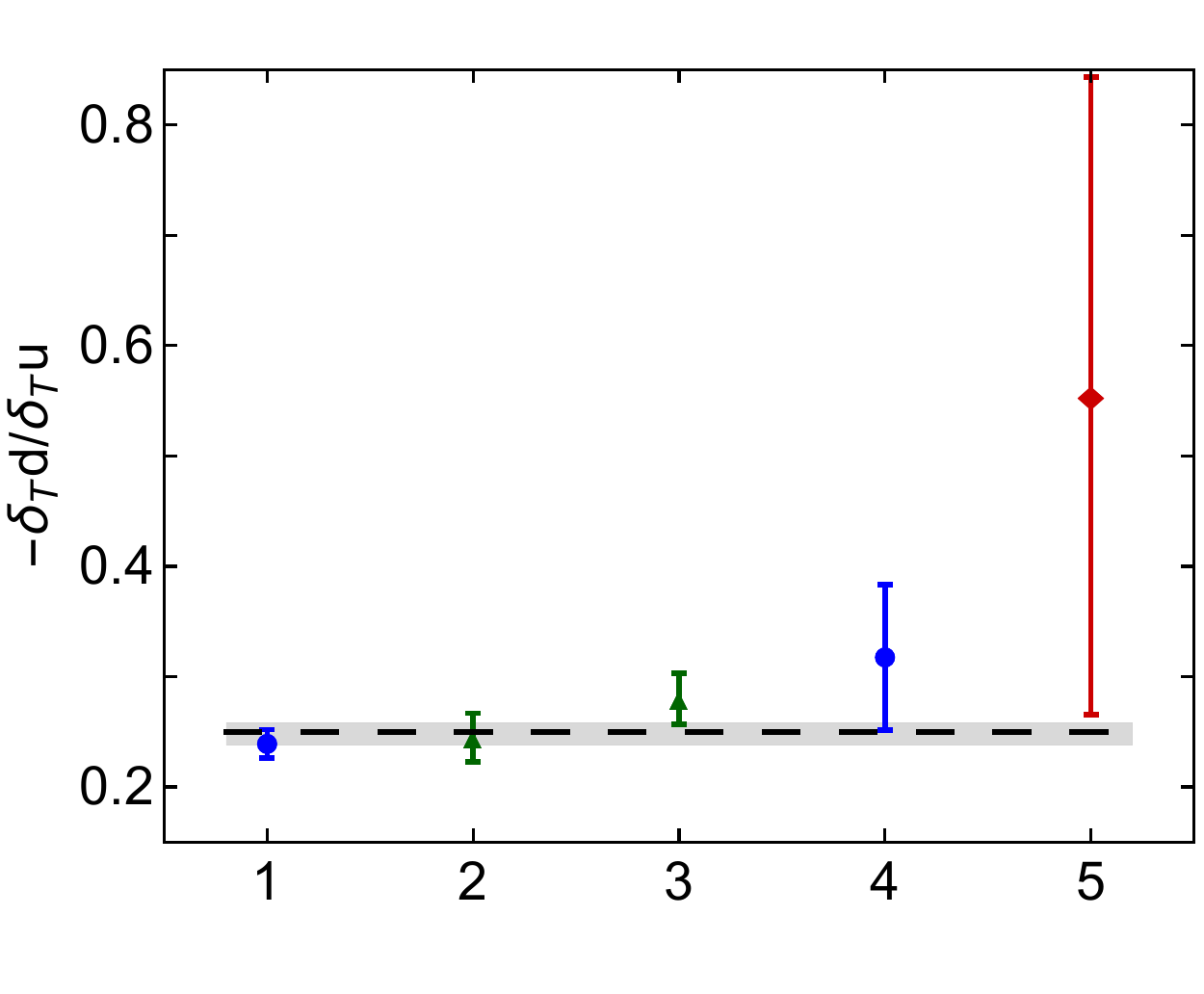}
\caption{\label{ratiocompare}
Ratio $(-\delta_T d/\delta_T u)$.  Position 1: our result;
lQCD: 2 -- \cite{Bhattacharya:2016zcn} and 3 -- \cite{Alexandrou:2017qyt};
contact-interaction Faddeev equation: 4 -- \cite{Xu:2015kta}.  The grey band depicts the weighted average in Eq.\,\eqref{DSElQCDratio}; and the dashed horizontal line is the quark model result $(-\delta_T d/\delta_T u)=1/4$ \cite{He:1994gz}.
Position 5 -- estimate obtained using Eq.\,\eqref{DefineTensorCharge} and extant transversity distribution data \cite{Ye:2016prn}.
}
\end{figure}

It is highlighted, too, by Fig.\,\ref{ratiocompare}, which depicts the scale-independent ratio $(-\delta_T d/\delta_T u)$.  In this case, the weighted average of theoretical predictions is
\begin{equation}
\label{DSElQCDratio}
-\overline{\frac{\delta_T d}{\delta_T u}} = 0.248(10)\,,
\end{equation}
illustrated by the grey band in the figure.  Using a simple nonrelativistic quark model spin-flavour wave function, this ratio is $0.25$.  It is practically the same in the MIT bag model \cite{He:1994gz}; but, in both cases, the individual tensor charges are measurably larger in magnitude than our results and those obtained using lattice methods \cite{Bhattacharya:2016zcn, Alexandrou:2017qyt}.

It is worth noting here that the nonrelativistic quark model prediction for the ratio of proton-to-neutron magnetic moments, $\mu_p/\mu_n$, agrees with the empirical value, but this outcome is also a numerical accident.  The quark model result is obtained neglecting meson-baryon final state interactions (meson cloud effects), which are known to contribute a roughly equal amount to $\mu_p$ and $\mu_n$ \cite{Cloet:2008wg}, increasing both in magnitude, so that the meson-undressed result should exceed the empirical value in size.

\section{Electric Dipole Moments}
\label{SecEDM}
Using Eqs.\,\eqref{EDMeqs} and \eqref{deltaTproton}, we have
\begin{equation}
\label{dndp}
\tilde d_p = 0.91\,\tilde d_u - 0.22\,\tilde d_d\,,\;
\tilde d_n = -0.22\,\tilde d_u + 0.91\,\tilde d_d\,.
\end{equation}
The impact of these results on beyond-SM phenomenology may be elucidated, \emph{e.g}.\ by following the analysis in Refs.\,\cite{Bhattacharya:2016zcn, Gao:2017ade}.

In this connection it is worth remarking that the possibility of a $s$-quark contribution produces some uncertainty in estimates of nucleon EDMs \cite{Chien:2015xha}, largely because its size is poorly known.  An estimate of this contribution is thus useful.  Such may be obtained via a simplified treatment of meson-loop corrections to the quark gap equations, as used elsewhere \cite{Cloet:2008fw, Chang:2009ae} to estimate the proton's strangeness-magnetic-moment and -$\sigma$-term.  Following that reasoning, we find $\delta_T s(\zeta_2) \approx 0.02 g_T^{0} = 0.014(1)$, a value consistent with contemporary lQCD estimates \cite{Bhattacharya:2016zcn, Alexandrou:2017qyt}.

\section{Epilogue}
\label{SecEpilogue}
We calculated the proton's tensor charges using the leading-order (rainbow-ladder, RL) truncation of all relevant matter-sector one-, two- and -three-valence-body equations, and the associated tensor current.  In particular, the three-body (Faddeev) equation is solved without recourse to a diquark approximation for the two-body scattering kernel.  Notably, once in possession of results for the tensor charges, one can use existing and future empirical limits on nucleon electric dipole moments to constrain extensions of the Standard Model.

Our results for the tensor charges [Eqs.\,\eqref{charges}] are commensurate with those obtained in contemporary lattice-QCD simulations [Figs.\,\ref{TCcompare} -- \ref{ratiocompare}].  This confluence increases tension between theory and phenomenology, \emph{viz}.\ whilst there is agreement on $\delta_T d$, direct computations of the tensor-charge matrix element [Eq.\,\eqref{tcd}] produce a value of $\delta_T u$ that is approximately twice as large as that obtained via analyses of extant transversity distribution data [Eq.\,\eqref{DefineTensorCharge}].  In a curious twist, the theoretical calculations produce a value of the scale-invariant ratio $(-\delta_T d/\delta_T u)$ which matches that obtained in simple quark models  [Fig.\,\ref{ratiocompare}], even though the individual charges are themselves very different.

This analysis completes the first improvement recommended in Ref.\,\cite{Xu:2015kta}, delivering continuum predictions of the tensor charges with a direct connection to QCD; and no material betterment of these results can be expected before methods are devised to improve over RL truncation in the three-body problem.

It may nevertheless be worth calculating the tensor charges using the QCD-kindred model employed successfully in computing a wide range of nucleon properties, including elastic and transition form factors \cite{Segovia:2015hra, Segovia:2015ufa, Segovia:2016zyc, Chen:2017pse, Roberts:2018hpf}.  As a model constrained by some data, elements of that framework implicitly improve upon RL truncation.  Hence, a comparison of its predictions with those presented herein could indicate what to expect from such refinements.

\acknowledgments
We are grateful for insights provided by L.~Chang, Z.-F.~Cui, G.~Eichmann, H.~Gao, J.-P.~Chen and J~ Segovia;
and for the hospitality of RWTH Aachen University, III.\,Physikalisches Institut B, Aachen, Germany.
Research supported by:
Forschungszentrum J\"ulich GmbH;
and U.S.\ Department of Energy, Office of Science, Office of Nuclear Physics, contract no.~DE-AC02-06CH11357.


\end{document}